\documentclass[twocolumn,amsmath,amssymb,pre,showpacs]{revtex4}

\usepackage{graphicx}
\usepackage{dcolumn}
\usepackage{bm}
\usepackage{amssymb}
\usepackage{amsmath}
\usepackage{amsthm,amsfonts}

\begin{document}
\bibliographystyle{aip}

\title{Rectification and One-Way Street for the Energy Current in Boundary-Driven Asymmetric Quantum Spin Chains}

\author{Emmanuel Pereira}
 \email{emmanuel@fisica.ufmg.br}
\affiliation{Departamento de F\'{\i}sica--Instituto de Ci\^encias Exatas, Universidade Federal de Minas Gerais, CP 702,
30.161-970 Belo Horizonte MG, Brazil}

\date{\today}

\begin{abstract}
Motivated by the demand of efficient quantum devices to engineer the energy transport, we analyze some  inhomogeneous quantum spin systems, including the
$XXZ$ chains, with magnetization baths at the ends. Aimed at finding general properties, we study the effects of
suitable transformations on the boundary-driven Lindblad master equation associated to the dynamics of the systems.  For  asymmetric models with target polarization at the edges or twisted XY boundary
gradients,  we show properties of the steady state which establish features of the energy current, irrespective of the system size and of the regime of
transport. We show the ubiquitous occurrence of energy rectification and, more
interestingly,  of an unusual phenomenon: in the absence of external magnetic field, there is an one-way street for the energy current, i.e., the direction of the
energy current does not change as we invert the magnetization baths at the boundaries. Given the extensiveness of the procedures, which essentially involve properties of the Lindblad master equation,
our results certainly follow for other interactions and
other boundary conditions. Moreover, our results indicate graded spin chains as genuine quantum rectifiers.
\end{abstract}


\def \Z {\mathbb{Z}}
\def \R {\mathbb{R}}
\def \La {\Lambda}
\def \la {\lambda}
\def \ck {l}
\def \F {\mathcal{F}}
\def \M {\mathcal{M}}
\newcommand {\md} [1] {\mid\!#1\!\mid}
\newcommand {\be} {\begin{equation}}
\newcommand {\ee} {\end{equation}}
\newcommand {\ben} {\begin{equation*}}
\newcommand {\een} {\end{equation*}}
\newcommand {\bg} {\begin{gather}}
\newcommand {\eg} {\end{gather}}
\newcommand {\ba} {\begin{align}}
\newcommand {\ea} {\end{align}}
\newcommand {\tit} [1] {``#1''}


\maketitle

\let\a=\alpha \let\b=\beta \let\d=\delta \let\e=\varepsilon
\let\f=\varphi \let\g=\gamma \let\h=\eta    \let\k=\kappa \let\l=\lambda
\let\m=\mu \let\n=\nu \let\o=\omega    \let\p=\pi \let\ph=\varphi
\let\r=\rho \let\s=\sigma \let\t=\tau \let\th=\vartheta
\let\y=\upsilon \let\x=\xi \let\z=\zeta
\let\D=\Delta \let\G=\Gamma \let\L=\Lambda \let\Th=\Theta
\let\P=\Pi \let\Ps=\Psi \let\Si=\Sigma \let\X=\Xi
\let\Y=\Upsilon

 The derivation of the laws of transport from the underlying microscopic models is one of the purposes of nonequilibrium statistical physics. Many works are devoted to this subject, and, nowadays,
 besides the study of fundamental questions, such as the onset of Fourier's law, a bedrock of heat transport theory,  we observe a considerable effort which aims at understanding the mechanisms of manipulation and control of the heat or energy current \cite{BLiRMP}, including experimental works \cite{Chang}. These studies are mainly inspired by the amazing progress of modern electronics due to the invention of electrical diodes, transistors and other nonlinear solid-state devices. The current lacking  of an efficient analog of diode, i.e., the lacking of a device which clearly has a preferential
direction for the energy flow \cite{CPC}, prevents the advance of phononics, the counterpart of electronics dedicated to the manipulation of the energy current, and makes several authors to devote their investigations to the mechanism of rectification.

Given  the present ambient of miniaturization, the possibility of quantum effects and the exiguity of quantum results in this specific area, we propose to investigate the mechanism of energy transport and rectification in genuine quantum models. In the present work, we consider the analysis of the energy (and spin) currents in some open quantum spin chains, including versions of the XXZ models. We emphasize that the detailed investigation of these
systems, such as the XXZ chain, the archetypal model of open quantum system, besides being profitable for the understanding of quantum effects on the mechanism
 of the energy current,  it also involves issues which interest to many areas, such as optics and cold-atoms, condensed matter, quantum information, etc. \cite{T}.

We analyze some one-dimensional spin models driven out equilibrium by magnetization reservoirs coupled to the boundary spins, i.e., by the presence of pumping applied at the edges. More
precisely, we analyze some spin models with dynamics given by boundary-driven Lindblad equations. We remark that such a choice of reservoirs does not precisely describe a quantum system passively coupled to heat baths \cite{FBarra} (problem to be treated in future works). But it is worth stressing that these specific quantum spin chains, e.g., the XXZ versions with magnetization pumping, are also of great theoretical interest \cite{Prosen+}. Moreover, these models can also be experimentally realized due to the advance of nanotechnology, which  allows us to manipulate different materials, even those with few elements, and to engineer different quantum reservoirs and specific designs for the coupling between systems and reservoirs \cite{Z}.

Inspired, in some way, by the work of Popkov and Livi \cite{PopLi}, our strategy is the following: instead of performing extensive and intricate computations to determine the steady density matrix, we simply study the action on the Lindblad master equation (LME)
of some properly chosen operators, related to the inversion of the baths, i.e., to the permutation between the bath linked to the first site and the bath linked to the last site. And so,
we determine the effects of such transformations on the stationary density matrix, as well as  on the energy and spin currents. As outcome, we reveal properties, independent of the system size and regime of transport, which lead to the occurrence of energy rectification and, more interestingly, to a particular phenomenon: in the absence of magnetic field, for the inhomogeneous, asymmetric (e.g., graded) XXZ model with target polarization at the edges or twisted XY boundary gradients, and certainly other boundary conditions and different interactions, there is an one-way street for the energy current. Namely, the direction of the energy current is determined by the asymmetry in the chain, not by the magnetization baths. In other words, not
only the magnitude, but also the direction of the energy flow does not change as we invert the magnetization baths at the boundaries.

We must stress that the existence of asymmetry in the structure of a given model is absolutely not a guarantee for the existence of energy rectification. The ingredients for asymmetry in the energy flow, i.e., for the occurrence of rectification and related effects, are not trivial. In order to make transparent that it is indeed an intricate problem, before describing our investigations on the spin chains, we briefly revisit the classical chain of oscillators, in which this problem of rectification has been already investigated in details.

Since the pioneering works of  Debye and Peierls, the prototype model for heat conduction in insulating solids is given by chains of anharmonic oscillators. Consequently, such systems
have been recurrently and exhaustively studied \cite{LLPDhar}, in particular, in reference to the energy rectification \cite{Casati+}.
For transparency, we will succinctly analyze the phenomenon in a minimal chain of 3 sites. We repeat that a more detailed study in larger systems is already known \cite{Prapid}.

As well known, Fourier's law holds in many systems: here, let us focus on chains of oscillators with nearest neighbor harmonic interparticle interactions and anharmonic onsite potentials \cite{Prapid}. In such cases, we usually have a thermal conductivity depending on the local temperature and other parameters. To investigate rectification in this chain of oscillators, we start from the expression for local Fourier's law derived for such systems in previous works. For inhomogeneous chains \cite{Prapid}, the Fourier's law in any part of the chain, i.e., the heat flow from site $j$ to $j+1$, $\mathcal{F}_{j,j+1}$, is given by
\begin{equation}
\mathcal{F}_{j,j+1} = -\kappa_{j,j+1}(\nabla T)_{j} = \frac{-1}{c_{j}T_{j}^{\alpha} +
{c}_{j+1}T_{j+1}^{\alpha}} \left( T_{j+1}-T_{j} \right) ,\label{fluxo}
\end{equation}
where $T_{j}$ is the local temperature, $\alpha \geq 0$, and $c_{j}$ depends on local parameters (particle mass, etc.). In a graded system, we have $c_{j-1} > c_{j} > c_{j+1}$ or $c_{j-1} < c_{j} < c_{j+1}$.

Hence, for a chain of 3 sites,
\begin{equation*}
\mathcal{F}_{1,2} = - \kappa_{1,2} (T_{2} - T_{1})~, \mathcal{F}_{2,3} = - \kappa_{2,3} (T_{3} - T_{2})~.
\end{equation*}
In the steady state, where $\mathcal{F}=\mathcal{F}_{1,2} = \mathcal{F}_{2,3}$, we have
\begin{equation*}
\mathcal{F} = \frac{-1}{c_{1}T_{1}^{\alpha} + c_{2}T_{2}^{\alpha}} (T_{2} - T_{1})
 = \frac{-1}{c_{2}T_{2}^{\alpha} + c_{3}T_{3}^{\alpha}} (T_{3} - T_{2}) .
\end{equation*}
And summing up the 2 parts of the equation above, we get
\begin{equation*}
\mathcal{F} = \frac{-1}{c_{1}T_{1}^{\alpha} + 2c_{2}T_{2}^{\alpha} + c_{3}T_{3}^{\alpha}} (T_{3} - T_{1})~.
\end{equation*}

To determine the heat flow, we recall that the temperatures $T_{1}$ and $T_{3}$ at the edges are given a priori, and so,  $T_{2}$ is computed by using the equations above.
The computation becomes very simple if we still assume a very small gradient of temperature in the chain, namely, $T_{1} = T + a_{1}\varepsilon$ and $T_{3} = T + a_{3}\varepsilon$
(again, given $T$, $a_{1}$, $a_{3}$ and $\varepsilon$, which is small). Hence, considering the result only up to $\mathcal{O}(\varepsilon)$, algebraic manipulations give us
\begin{equation*}
a_{2} = \frac{1}{2c_{2} + c_{1} + c_{3}} \left[ c_{1}a_{3} + c_{2}(a_{1} + a_{3}) + c_{3}a_{1}\right] .
\end{equation*}
And so, the temperature $T_{2}$ and the heat flow $\mathcal{F}$ are determined. Note that we may write
\begin{equation*}
\mathcal{F} = -\kappa(T_{3} - T_{1}) ~~ \Rightarrow ~~ \kappa = 1/\left( c_{1}T_{1}^{\alpha} + 2c_{2}T_{2}^{\alpha} + c_{3}T_{3}^{\alpha} \right) ~.
\end{equation*}

Carrying out the computation for the inverted system, i.e., for the chain in which $a'_{1} = a_{3}$ and $a'_{3} = a_{1}$, we obtain $a'_{2}$, which will be given by the expression for
$a_{2}$ above but with the replacement $a_{1} \leftrightarrow a_{3}$,
\begin{equation*}
a'_{2} = \frac{1}{2c_{2} + c_{1} + c_{3}} \left[ c_{1}a_{1} + c_{2}(a_{1} + a_{3}) + c_{3}a_{3}\right] ~.
\end{equation*}
 Moreover, we obtain the inverted flow $\mathcal{F}' = -\kappa'(T_{1} - T_{3})$, where $\kappa'$ is such that
\begin{equation} \label{kappas}
\frac{1}{\kappa} - \frac{1}{\kappa'} = \alpha\varepsilon T^{\alpha - 1} (c_{1} - c_{3})(a_{1} - a_{3}) \left[\frac{ c_{1} +c_{3}}{2c_{2}  + c_{1} +c_{3}}\right] ~.
\end{equation}

  Despite the simplicity of this minimal model, important information can be extracted from the equations above. For inhomogeneous (graded) chains and generic temperatures at the edges, an interesting feature is that the temperature profile $T'$, for the system with inverted baths ($T'_{1} = T_{3}$ and $T'_{3} = T_{1}$), is different from that obtained by inverting the original temperature profile $T$. That is, if we simply invert the original
 profile, we get  $a'_{1} = a_{3}$, $a'_{3}= a_{1}$ and
$a'_{2} = a_{2}$, which is not the profile for the inverted chain. And it follows even for the specific case of $\alpha = 0$, that describes a system in which the thermal conductivity does not depend on temperature.
 By the other side, rectification holds only if $\alpha \neq 0$, see Eq.(\ref{kappas}).

In other words, if we invert an asymmetric chain of oscillators between two baths, even if we observe a different temperature profile, i.e.,  a final profile which is not the inversion of the original one, the occurrence of
rectification is not mandatory. In short, asymmetry in a chain is not a synonym of rectification, i.e., it is not a synonym of asymmetry in the energy current \cite{Note,Psuf+}.

A further comment is relevant. It is rigorously proved \cite{PLA} that rectification is absent in any asymmetric version of the harmonic chain of oscillators with self-consistent inner stochastic baths - these inner baths mimic the anharmonic potentials absent in the initial Hamiltonian. Such a model obeys the Fourier's law as described above, with $\alpha= 0$. In short, the previous derived result is not due to the size of the chain, nor due to the approximation in the temperature gradient.

Now, we introduce the spin systems.

We consider here the one-dimensional quantum  spin model, with Hamiltonian (for $\hbar = 1$)
\begin{eqnarray}
\mathcal{H} = && \sum_{i=1}^{N-1}\left\{ \alpha_{i,i+1}\sigma_{i}^{x}\sigma_{i+1}^{x} + \alpha'_{i,i+1}\sigma_{i}^{y}\sigma_{i+1}^{y}  + \Delta_{i,i+1}\sigma_{i}^{z}\sigma_{i+1}^{z} \right\}\nonumber \\
 && + \sum_{i=1}^{N} B_{i}\sigma_{i}^{z} ~, \label{hamiltonian}
\end{eqnarray}
where $\sigma_{i}^{\beta}$ ($\beta = x, y, z$) are the Pauli matrices and $B_{i}$ is the external magnetic field acting on site (particle) $i$. In fact, we can consider several other $\mathcal{H}$: the main restriction is the invariance of $\mathcal{H}$ under the transformations to be presented. Anyway, for simplicity, we will restrict the analysis to the simple case of the XXZ version with $\alpha_{i,i+1} = \alpha'_{i,i+1} = \alpha$.

The time evolution
of the system density matrix $\rho$ is given by a Lindblad quantum master equation \cite{BP}
\begin{equation}
\frac{d\rho}{d t} = i[\rho, \mathcal{H}] + \mathcal{L}(\rho) ~.\label{master}
\end{equation}
The dissipator $\mathcal{L}(\rho)$ describes the coupling with the baths, and it is given by
\begin{eqnarray}
\mathcal{L}(\rho) &=& \mathcal{L}_{L}(\rho) + \mathcal{L}_{R}(\rho) ~, \nonumber\\
\mathcal{L}_{L,R}(\rho) &=& \sum_{s=\pm} L_{s}\rho L_{s}^{\dagger} -
\frac{1}{2}\left\{ L_{s}^{\dagger}L_{s} , \rho \right\} ~.\label{dissipator}
\end{eqnarray}

For $\mathcal{L}_{L}$, in the case of a XXZ chain with target $\sigma^z$ polarization at the edges,  we have
\begin{equation}
L_{\pm} = \sqrt{\frac{\gamma}{2}(1 \pm f_{L})} \sigma_{1}^{\pm} ~\label{dissipator2},
\end{equation}
and similarly  for $\mathcal{L}_{R}$, but with $\sigma_{N}^{\pm}$ and $f_{R}$ replacing  $\sigma_{1}^{\pm}$ and $f_{L}$. In the previous expressions, $\{\cdot,\cdot\}$ denotes the anticommutator;
$\sigma_{j}^{\pm}$ are the spin creation and annihilation operators $\sigma_{j}^{\pm} = (\sigma_{j}^{x} \pm i\sigma_{j}^{y})/2$~; $\gamma$ is the coupling strength to the spin baths; $f_{L}$ and $f_{R}$ give the
driving strength, and are related to the polarization of extra spin  at the boundaries: $f_{L} = \langle\sigma_{0}^{z}\rangle$ and  $f_{R} = \langle\sigma_{N+1}^{z}\rangle$.

The expressions for spin and energy currents follow from the LME and continuity equations, see e.g. Ref.\cite{Mendoza-A} for detailed derivations. At site $j$, inside the chain, for the magnetization current $\langle J_{j}\rangle$ and energy current $\langle F_{j} \rangle$, we have
\begin{eqnarray}
\langle J_{j} \rangle &=& 2\alpha \langle \sigma_{j}^{x} \sigma_{j+1}^{y} - \sigma_{j}^{y}\sigma_{j+1}^{x} \rangle ~,  \nonumber \\
\langle F_{j}\rangle &=& \langle F_{j}^{XXZ}\rangle + \langle F_{j}^{B}\rangle ~,\nonumber\\
\langle F_{j}^{XXZ} \rangle &=&  2\alpha \langle \alpha \left( \sigma_{j-1}^{y}\sigma_{j}^{z} \sigma_{j+1}^{x} - \sigma_{j-1}^{x}\sigma_{j}^{z}\sigma_{j+1}^{y}\right) \nonumber\\
&& + \Delta_{j-1,j}\left( \sigma_{j-1}^{z}\sigma_{j}^{x} \sigma_{j+1}^{y} - \sigma_{j-1}^{z}\sigma_{j}^{y}\sigma_{j+1}^{x}\right) \nonumber\\
&& + \Delta_{j,j+1}\left( \sigma_{j-1}^{x}\sigma_{j}^{y} \sigma_{j+1}^{z} - \sigma_{j-1}^{y}\sigma_{j}^{x}\sigma_{j+1}^{z}\right)\rangle  ~,\nonumber\\
\langle F_{j}^{B} \rangle &=& \frac{1}{2} B_{j}\langle J_{j-1} + J_{j}\rangle ~. \label{currents}
\end{eqnarray}

To carry out the study,
 we take $f = f_{L} = -f_{R}$. And so, within such a choice, the inversion of the baths at the edges ($f_{L} \leftrightarrow f_{R}$) is given by the change in the sign of $f$.

Analyzing the dissipator $\mathcal{L}(\rho)$, we note that it does not modifies if we change the sign of $f$ and, at the same time, make the replacements $\sigma_{1}^{+} \leftrightarrow \sigma_{1}^{-}$ and
$\sigma_{N}^{+} \leftrightarrow \sigma_{N}^{-}$.

Hence, in the case of absence of the external magnetic field, $B\equiv 0$, we investigate the effects of the following transformation [the idea is to find a suitable transformation, related to the inversion, i.e., permutation of the baths]
\begin{equation}
U = \sigma_{1}^{x}\otimes\sigma_{2}^{x}\otimes \ldots \otimes\sigma_{N}^{x} ~, U^{\dagger} = U^{-1} ~.
\end{equation}
 From the LME we have
\begin{eqnarray*}
\lefteqn{\frac{d}{dt}\left[ U^{-1}\rho U\right] = iU^{-1}\rho\mathcal{H}U - iU^{-1}\mathcal{H}\rho U + U^{-1}\mathcal{L}U} \\
 &&+ iU^{-1}\rho UU^{-1}\mathcal{H}U - iU^{-1}\mathcal{H}UU^{-1}\rho U + U^{-1}\mathcal{L}U ~.
 \end{eqnarray*}
 But, $U^{-1}\mathcal{H}U = \mathcal{H}$ (for $B=0$), and $U^{-1}\mathcal{L}(f)U = \mathcal{L}(-f)$. That is, if $\rho$ is a solution of the LME, then $U^{-1}\rho U$ is a solution of the LME with $-f$, which means, it is a solution of
 the system with inverted baths. Recalling the uniqueness of the stationary solution of these LME \cite{PopLi, Pop1, EvansProsen}, we can say that if $\rho$ is the steady solution of the LME, then $U^{-1}\rho U$ is the
 steady solution of the LME with $-f$, i.e., a solution of the system with inverted baths.

In relation to the effect of $U$ on the currents, we have
\begin{equation}
U^{-1}F^{XXZ}_{j}U = F^{XXZ}_{j}~,
\end{equation}
see Eqs.(\ref{currents}). Consequently,
\begin{eqnarray*}
\langle F^{XXZ}_{j} \rangle &\equiv & tr(\rho F^{XXZ}_{j}) = tr(\rho U^{-1}F^{XXZ}_{j}U) \\
&= & tr(U^{-1}\rho U F^{XXZ}_{j}) = tr(\rho(-f) F^{XXZ}_{j}) ~.
\end{eqnarray*}
Precisely, $\langle F^{XXZ}_{j} \rangle = \langle F^{XXZ}_{j} \rangle_{ib}$, where $ib$ means the system with inverted baths.
In other words, $\langle F^{XXZ}_{j}(f) \rangle = \langle F^{XXZ}_{j}(-f) \rangle$, i.e., $\langle F^{XXZ}_{j} \rangle$ is an even function of $f$.

The implications of such property are evident.
First, for the case of a homogeneous chain, it proves the vanishing of  $\langle F^{XXZ}_{j} \rangle$, as described in Ref.\cite{PopLi}, there with the use of
the left-right reflexion operation. To show the vanishing here, note that in the homogeneous chain, there is nothing  which may indicate the direction for the energy flow - it must be given by the baths. And so, if we invert the baths, the direction of the energy flow must invert. As it
does not happen due to the property above, we must conclude that $\langle F^{XXZ}_{j} \rangle$ vanishes in the homogeneous system. And, for the case of a graded chain (note that, in such case, the reflexion operation does not work anymore), in any situation which allows an energy current, its direction is determined by some structure in the chain, not by the baths. It means that we have a one-way street for the energy flow, say, a ``complete, integral rectification''.

And, for the magnetization current $J_{j}$,
\begin{equation}
U^{-1} J_{j}U = - J_{j}~,
\end{equation}
see Eq.(\ref{currents}).  Then,
\begin{eqnarray*}
\langle J_{j} \rangle &\equiv & tr(\rho J_{j}) = - tr(\rho U^{-1}J_{j}U) \\
&= & - tr(U^{-1}\rho U J_{j}) = - tr(\rho(-f) J_{j}) ~.
\end{eqnarray*}
In short, $\langle J_{j}(f) \rangle = - \langle J_{j}(-f) \rangle$, namely, $\langle J_{j} \rangle$ is an odd function of $f$. It means that the direction of the magnetization flow is given by the baths: if we invert the baths,
the direction of the flow is inverted. Note also that rectification is absent: the magnitude of the spin current is preserved.

The extension of our findings for the system in the presence of a homogeneous magnetic field $B$ is immediate. Indeed, a homogeneous $B$ does not affect expectation values of spin-conserving operators (more comments and
details in Ref.\cite{Mendoza-A} and references there in), and so, both $\langle F^{XXZ}_{j} \rangle$ and $\langle J_{j} \rangle$ do not change. Now, the total energy current is given by, see Eqs.(\ref{currents}),
\begin{equation}
\langle F_{j} \rangle = \langle F^{XXZ}_{j} \rangle + B\langle J_{j} \rangle ~,
\end{equation}
namely, it is a sum of an even function of $f$ with another odd function of $f$. Consequently, unless vanishing, $\langle F(f) \rangle \neq \langle F(-f) \rangle$, that is, the occurrence of energy rectification is transparent.
Note also that the presence of a large magnetic field $B$ decreases the rectification power.

It is important to report that, in a recent work \cite{SPL}, considering the investigation of energy transport in the open quantum XXZ chain, we perform analytical computations for the energy and spin currents by finding the stationary density matrix of the associated LME. Precise algebraic formulas are obtained for a minimal chain with 3 sites (the restriction to small systems was due to technical difficulties), and numerical computations extend some results to systems up to 8 sites. An exact and huge expression for $\langle F \rangle$ is derived in Ref.\cite{SPL} for the minimal chain. For clearness, we write below the dominant terms in a expansion in powers of $f$, the driving strength, and of $\delta$, the asymmetry parameter, defined ahead. For the Hamiltonian (\ref{hamiltonian}) of the graded chain with 3 sites, we take $\alpha_{i,i+1} = \alpha'_{i,i+1} = 1$, $\Delta_{1,2} = \Delta - \delta$, $\Delta_{2,3} = \Delta + \delta$, $B_{i} = B$; and so, we have
\begin{eqnarray*}
\langle F \rangle &=& Bf \left( \frac{912}{969 + 48\Delta^{2}}\right) \\
 && + f^{2}\delta \left( \frac{32(20224\Delta^{4} + 64256\Delta^{2} -1083)}{(51 + 16\Delta^{2})(323 + 16\Delta^{2})^{2}}\right) ~.
\end{eqnarray*}
The formula above for the
energy current, in the minimal chain, shows that it is nonzero, for $f \neq 0$, even if $B=0$. Moreover, again for $B=0$, the energy current does not change as we invert the baths. These results follow also in the exact formula (beyond $\mathcal{O}(f^2)$). Our analysis, carried out in the present work by means of a completely different approach, shows that the
energy rectification and the one-way street for energy current as $B=0$
are  not only an accident in a very small chain, but  they are features of these graded XXZ systems and many other spin models with the same symmetries.

Another relevant version of the XXZ model is given by the chain coupled to boundary baths which tend to polarize the spins at the ends along different directions \cite{Pop1}. Now, beginning with $B= 0$, we analyze the version
 in which the dissipators in the  LME are given by
 \begin{eqnarray}
\mathcal{L}_{L}(\rho) &=& -\frac{1}{2} \sum_{m=1}^{2} \left\{ \rho, W^{\dagger}_{m}W_{m} \right\} + \sum_{m=1}^{2} W_{m} \rho W^{\dagger}_{m} ~, \nonumber\\
\mathcal{L}_{R}(\rho) &=& -\frac{1}{2} \sum_{m=1}^{2} \left\{ \rho, V^{\dagger}_{m}V_{m} \right\} + \sum_{m=1}^{2} V_{m} \rho V^{\dagger}_{m} ~,
\end{eqnarray}
with
\begin{eqnarray*}
W_{1} &=& \sqrt{\frac{1-k}{2}} \left( \sigma^{z}_{1} + i\sigma_{1}^{x} \right) ~,  W_{2} = \sqrt{\frac{1+k}{2}} \left( \sigma^{z}_{1} - i\sigma_{1}^{x} \right) ~, \\
V_{1} &=& \sqrt{\frac{1+k'}{2}} \left( \sigma^{y}_{N} + i\sigma_{N}^{z} \right) ~,  V_{2} = \sqrt{\frac{1-k'}{2}} \left( \sigma^{y}_{N} - i\sigma_{N}^{z} \right) ~,
\end{eqnarray*}
where we will take $-1 \leq k \leq 1$, and, in our case, $k'=-k$.

To carry out the analysis, we define the operator $\sigma^{r}$, where
\begin{eqnarray*}
\sigma^{r} \equiv  \left(\begin{array}{cc} 0 & 1 \\
 i & 0 \end{array} \right)
~~\Rightarrow \left(\sigma^{r}\right)^{\dagger} =  \left(\sigma^{r}\right)^{-1} =  \left(\begin{array}{cc} 0 & -i \\
 1 & 0 \end{array} \right) ~.
\end{eqnarray*}
[Of course, in order to obtain a suitable transformation $\sigma^{r}$, we have performed a detailed algebraic study involving the effects of generic transformations on
the terms of the LME.]

It follows that
\begin{equation*}
\sigma^{r} \sigma^{x} \left(\sigma^{r}\right)^{\dagger} = \sigma^{y} ~, ~~ \sigma^{r} \sigma^{y} \left(\sigma^{r}\right)^{\dagger} = \sigma^{x} ~,~~
\sigma^{r} \sigma^{z} \left(\sigma^{r}\right)^{\dagger} = -\sigma^{z} ~.
\end{equation*}

Discarding the indices $1$ and $N$ in $\sigma^{x}, \sigma^{y}, \sigma^{z}$ written in $W_{1}, W_{2}, V_{1}$ and $V_{2}$, we have
\begin{eqnarray*}
\sigma^{r} W_{1} (\sigma^{r})^{\dagger} &=& iV_{1} ~, ~~\sigma^{r} W_{2} (\sigma^{r})^{\dagger} = -iV_{2} ~,\\
\sigma^{r} V_{1} (\sigma^{r})^{\dagger} &=& -iW_{1} ~, ~~\sigma^{r} V_{2} (\sigma^{r})^{\dagger} = iW_{2} ~, ~~ {\rm etc.}
\end{eqnarray*}

Thus, for the analysis of (a)symmetries in the LME, we define
\begin{equation}
U^{\dagger} = \sigma_{1}^{r}\otimes\sigma_{2}^{r}\otimes \ldots \otimes\sigma_{N}^{r} ~.
\end{equation}

And we have that, if $\rho$ is a solution of the LME, then $U^{\dagger}\rho U$ is a solution of the LME, but with inverted baths.

Turning to the currents, we have
\begin{equation*}
U^{\dagger} F_{j}^{XXZ} U = F_{j}^{XXZ} ~, ~~ U^{\dagger} J_{j} U = -J_{j} ~.
\end{equation*}
And so, it follows an analysis similar to the previous one, presented for the case of target spins at the edges of the chain: $\langle F^{XXZ}_{j} \rangle = \langle F^{XXZ}_{j} \rangle_{ib}$, $\langle J_{j} \rangle = - \langle J_{j} \rangle_{ib}$, etc. Consequently, we find similar phenomena: one-way street  and rectification.

To conclude, we make some remarks.

In Ref.\cite{PopLi}, the authors analyze homogeneous XXZ driven spin chains by considering symmetries of the LME. They show that different pumping applied at the edges, irrespective of
the system size or the regime of transport, can switch on and off the spin and/or energy currents in the stationary state. In the present work, motivated by
the investigation of the energy rectification phenomenon (evidently, in asymmetric models), we also study the behavior of the LME under suitable transformations. We show that, again, irrespective of the system size or the regime of transport, in inhomogeneous, asymmetric spin chains we can find energy rectification or the emergence of
an one-way street for the energy current, even in a chain with no energy current in the homogeneous case.

Some interesting recent results concerning the rectification of the spin current by direct computation in the steady state are presented in Ref.\cite{Prosen2}. There, the authors consider  XXZ chains with target $\sigma^z$ polarization at the edges,  and in the presence of an inhomogeneous external magnetic field. Another interesting and recent result concerning transport in spin chains is presented in Ref.\cite{Z}, where the authors find
diffusive and subdiffusive high temperature spin transport in a disordered Heisenberg chain in the ergodic regime.

A further comment on the interest of asymmetric models is pertinent. For instance, graded materials, i.e., systems whose structure or composition
changes gradually in space, are not only an academic issue. They are abundant in nature, can be experimentally manipulated, and have attracted attention in different areas, with works devoted to the investigation of their mechanical, electrical, optical, and heat conduction properties \cite{Graded}.

Finally, we stress that the present results indicate graded spin chains as quantum materials suitable for the building of rectifiers and other devices devoted to the manipulation of the energy current. We believe that they will
motivate more research on this topic.


\vspace*{1 cm} {\bf Acknowledgments:} This work was partially supported by CNPq (Brazil).


\begin{thebibliography}{32}
\bibitem{Note} For a graded chain in the regime of normal transport,  it is shown that the existence of a local thermal conductivity dependent on temperature and other
local parameters engenders sufficient conditions to guarantee the occurrence of thermal rectification \cite{Psuf+}.
\end{thebibliography}
\end{document}